%% file: ms.tex
\documentclass[a4paper]{article}

\usepackage{itgspeech2023}    %
\usepackage{times}            %
\usepackage[english]{babel}   %
\usepackage[ansinew]{inputenc}%
\usepackage[T1]{fontenc}      %
\usepackage{amsmath,amssymb}
\usepackage{graphicx}
\usepackage[colorlinks=false,pdfborder={0 0 0}]{hyperref}
\usepackage{units}

\usepackage{color}
\usepackage{amsfonts,physics}
\usepackage{algorithm}
\usepackage{array}
\usepackage{textcomp}
\usepackage{stfloats}
\usepackage{float}
\usepackage{verbatim}
\usepackage{setspace}
\usepackage[noend]{algpseudocode}

\usepackage{caption}
\usepackage{subcaption}
\usepackage{xcolor}

\usepackage{comment}
\usepackage{cite}
\usepackage{soul}
\usepackage{epstopdf}
\usepackage{xfrac}
\usepackage{pifont}

\usepackage{cleveref}
\usepackage{csquotes}
\usepackage{booktabs}
\usepackage{multirow}
\usepackage{tabularx,ragged2e}
\newcolumntype{Y}{>{\centering\arraybackslash}X}
\usepackage{footmisc}

\usepackage{varwidth}
\usepackage{tikz}
\usepackage{pgf}
\usepackage{pgfplots}
\usepackage{pgfplotstable}
\pgfplotsset{compat=1.3}
\usetikzlibrary{shapes.geometric}
\usetikzlibrary{shapes.arrows}
\usepackage{array}
\usepackage[nolist, nohyperlinks]{acronym}

\newcommand{\D}[1]{\mathrm{d}{#1}}
\newcommand{\state}[0]{\mathbf{x}_\tau}

\newcommand{\reverb}[0]{\mathbf{y}}
\newcommand{\sco}[0]{\nabla_{\state} \log p(\state)}
\newcommand{\smax}[0]{\sigma_\mathrm{max}}
\newcommand{\smin}[0]{\sigma_\mathrm{min}}

\title{Wind Noise Reduction with a Diffusion-based Stochastic Regeneration Model}

\author{Jean-Marie Lemercier,$^{\star}$
      Joachim Thiemann,$^{\dagger}$
      Raphael Koning,$^{\dagger}$
      Timo Gerkmann$^{\star}$}
\address{$^\star$ \{first.name\}@uni-hamburg.de, Universit{\"a}t                Hamburg, Hamburg, Germany \\
        $^\dagger$ \{first.name\}@advancedbionics.com, Advanced Bionics, Hannover, Germany
}

\begin{document}

\begin{acronym}
\acro{stft}[STFT]{short-time Fourier transform}
\acro{istft}[iSTFT]{inverse short-time Fourier transform}
\acro{dnn}[DNN]{deep neural network}
\acro{pesq}[PESQ]{Perceptual Evaluation of Speech Quality}
\acro{polqa}[POLQA]{perceptual objectve listening quality analysis}
\acro{wpe}[WPE]{weighted prediction error}
\acro{psd}[PSD]{power spectral density}
\acro{rir}[RIR]{room impulse response}
\acro{snr}[SNR]{signal-to-noise ratio}
\acro{lstm}[LSTM]{long short-term memory}
\acro{polqa}[POLQA]{Perceptual Objectve Listening Quality Analysis}
\acro{sdr}[SDR]{signal-to-distortion ratio}
\acro{estoi}[ESTOI]{Extended Short-Term Objective Intelligibility}
\acro{elr}[ELR]{early-to-late reverberation ratio}
\acro{tcn}[TCN]{temporal convolutional network}
\acro{rls}[RLS]{recursive least squares}
\acro{asr}[ASR]{automatic speech recognition}
\acro{ha}[HA]{hearing aid}
\acro{ci}[CI]{cochlear implant}
\acro{mac}[MAC]{multiply-and-accumulate}
\acro{vae}[VAE]{variational auto-encoder}
\acro{gan}[GAN]{generative adversarial network}
\acro{tf}[T-F]{time-frequency}
\acro{sde}[SDE]{stochastic differential equation}
\acro{ode}[ODE]{ordinary differential equation}
\acro{drr}[DRR]{direct to reverberant ratio}
\acro{lsd}[LSD]{log spectral distance}
\acro{sisdr}[SI-SDR]{scale-invariant signal to distortion ratio}
\acro{mos}[MOS]{mean opinion score}
\acro{map}[MAP]{maximum a posteriori}
\acro{rtf}[RTF]{real-time factor}
\end{acronym}

\maketitle

\begin{abstract}
\input{sections/abstract}
\end{abstract}

\section{Introduction}
\input{sections/intro}

\section{Diffusion-based generative models}
\input{sections/sgm}

\section{Stochastic regeneration model}
\input{sections/storm}

\section{Experimental Setup}
\input{sections/exp}

\section{Results and Discussion}
\input{sections/results}

\section{Conclusions and Future Work}
\input{sections/conclusion}

\small
\bibliographystyle{ieeetr}
\bibliography{refs23}

\end{document}

%% file: sections/abstract.tex
In this paper we present a method for single-channel wind noise reduction using our previously proposed diffusion-based stochastic regeneration model combining predictive and generative modelling. We introduce a non-additive speech in noise model to account for the non-linear deformation of the membrane caused by the wind flow and possible clipping. We show that our stochastic regeneration model outperforms other neural-network-based wind noise reduction methods as well as purely predictive and generative models, on a dataset using simulated and real-recorded wind noise.
We further show that the proposed method generalizes well by testing on an unseen dataset with real-recorded wind noise.
Audio samples, data generation scripts and code for the proposed methods can be found online\footnote{https://uhh.de/inf-sp-storm-wind}\footnote{This work has been funded by the Federal Ministry for Economic Affairs and Climate Action, project 01MK20012S, AP380. The authors are responsible for the content of this paper.}.

%% file: sections/intro.tex
Wind noise captured in microphone signals
is an important factor of intelligibility and quality loss in speech communications, and occurs for virtually all outdoor scenarios. 
Hearing-device users particularly suffer from wind noise presence, more than from other noise types \cite{Kochkin2010ConsumerSatisfaction}.
Wind acoustics are highly non-stationary, especially in case of strong wind as it adopts a turbulent behaviour close to microphones.
Furthermore, the corruption caused by wind noise exhibit
non-linear behaviours, due to the displacement of the microphone membrane by the air flow and saturation for high wind noise levels \cite{Nelke2016WindNR}.
Such non-stationarity and non-linearities
make enhancing speech corrupted with wind noise a very difficult challenge \cite{Nelke2016WindNR, Zakis2011}. 

Several traditional enhancement solutions leverage mu\-lti-channel processing \cite{Thuene2016MaximumLikelihoodAT} and often exploit the spatial coherence structure across microphones shaped by the local turbulent flow \cite{Nelke2016WindNR, 
Mirabilii2020SpatialCM, Franz2010MultichannelAF}.
Single-channel solutions include adaptive post-filtering \cite{Nemer2009SinglemicrophoneWN} and spectral enhancement exploiting the particular spectrum of wind noise \cite{
Nelke2014SingleCentroids 
}. Other approaches were designed using the fact that wind noise resides mostly in low-frequency regions. These methods discard the polluted low-frequency speech information, and aim to recreate a clean version of it based on artificial bandwidth extension or synthesis techniques \cite{Nelke2012BWE, Nelke2015Reconstruction}.

More recently, machine learning solutions were proposed \cite{Bai2018DNNWind, 
Lee2017DeepBL}, mostly relying on supervised predictive \linebreak 
learning, i.e. recovering clean speech from noisy speech based on a mapping learnt by a \ac{dnn} during training.
Generative models are a different class of machine learning techniques that learn a parameterization of the clean speech distribution and allow to generate multiple valid estimates instead of a single best estimate as for predictive approaches \cite{MurphyBook2}. 
Such generative methods include \acp{vae},
normalizing flows,
\acp{gan}
and diffusion models \cite{sohl2015deep
}.
Diffusion models were recently proposed for speech restoration tasks such as enhancement, dereverberation and bandwidth extension \cite{lu2022conditional, Welker2022SGMSE, Richter2022SGMSE++, Lemercier2022analysing}. Originally intended for image generation, they showed impressive results on speech restoration, notably outperforming their predictive counterparts on speech quality \cite{Lemercier2022analysing}. 
In previous work \cite{Lemercier2022storm}, we proposed to combine predictive and generative modelling to leverage both the fast inference and interference removal power of predictive approaches, and the sample quality and generalization abilities of generative models.
The resulting model was evaluated on additive noise and dereverberation separately. 

We aim here to investigate the performance of the proposed model for wind noise reduction.
We introduce a signal model approximation for speech in wind noise taking into account possible non-linearities such as membrane displacement and clipping which often occur for strong winds \cite{Nelke2016WindNR}.
We show that our stochastic regeneration model is able to highly increase the quality and intelligibility of speech in wind noise. We compare to DNN-based baselines for wind noise reduction, as well as purely generative and predictive models using the same DNN architecture as the proposed method.
We validate our algorithm on both the matched test split of our simulated dataset and an unseen speech in wind noise dataset using real-recorded wind noise samples.

%% file: sections/sgm.tex
Diffusion models are a class of generative models that iteratively generate data from noise based on a stochastic process parameterization \cite{sohl2015deep, 
song2021sde}. More specifically, they use a forward diffusion process during training to progressively degrade clean data with Gaussian noise and/or other types of corruption. At inference time, a reversed version of the diffusion process generates a sample from the target data distribution given an initial Gaussian noise state.

\subsection{Forward and reverse processes}\label{sec:processes}

The stochastic forward process $\{\mathbf x_\tau\}_{\tau=0}^T$ is defined as a \ac{sde} \cite{
song2021sde}:
    \begin{equation} \label{eq:forward-sde}
        \D{\mathbf x_\tau} = \mathbf{f}(\mathbf x_\tau, \tau) \D{\tau} + g(\tau) \D{\mathbf w},
    \end{equation}
where $\mathbf x_\tau$ is the current state of the process indexed by the continuous time step $\tau \in [0, T]$. This \textit{diffusion time} variable $\tau$ relates to the progress of the stochastic process and should not be mistaken for our usual notion of time in time-series-like signals. The initial condition represents target clean speech $\mathbf{x}_0 = \mathbf{x}$. As our process is defined in the complex spectrogram domain, independently for each \ac{tf} bin, the variables in bold are assumed to be vectors in $\mathbb C^d$ containing the coefficients of the complex spectrogram--- with $d$ the product of the time and frequency dimensions--- whereas variables in regular font represent real scalar values. 
The stochastic process $\mathbf w$ is a standard $d$-dimensional Brownian motion, that is, $\D{\mathbf w}$ is a zero-mean Gaussian random variable with standard deviation $\D{\tau}$ for each \ac{tf} bin.
The \emph{drift} function $\mathbf f$ and \emph{diffusion} coefficient $g$ as well as the initial condition $\mathbf{x}_0$ and the final diffusion time $T$ uniquely define the 
process $\{\mathbf x_\tau\}_{\tau=0}^T$
Under some regularity conditions on $\mathbf f$ and $g$%
, the reverse process $\{\mathbf x_\tau\}_{\tau=T}^0$ is another diffusion process and is also the solution of a \ac{sde} \cite{anderson1982reverse, song2021sde}:
\begin{equation}\label{eq:reverse-sde}
        \resizebox{0.9\hsize}{!}{%
        $\D{\mathbf x_\tau} = \left[
            -\mathbf f(\mathbf x_\tau, \tau) + g(\tau)^2\nabla_{\mathbf x_\tau} \log  p_\tau(\mathbf x_\tau)
        \right] \D{\tau}
        + g(\tau)\D{\bar{\mathbf w}},$
        }
\end{equation}
where $\D{\bar{\mathbf w}}$ is a $d$-dimensional Brownian motion for the time flowing in reverse and $\nabla_{\mathbf x_\tau} \log p_\tau(\mathbf x_\tau)$ is the \emph{score function}, i.e. the gradient of the logarithmic data distribution for the current process state $\mathbf x_\tau$.

In order to perform speech restoration
, the generation of clean speech $\mathbf{x}$ is conditioned on cues depending on the noisy speech $\mathbf{y}$.
Previous diffusion-based approaches proposed to condition the process explicitly within the neural network \cite{chen2021wavegrad} or through guided classification \cite{dhariwal2021diffusion}. 
In \cite{Welker2022SGMSE} however, 
it has been proposed to include the conditioning information directly into the diffusion process by defining the forward process as the solution to the following Ornstein-Uhlenbeck \ac{sde}:
\begin{equation}\label{eq:ouve-sde}
        \resizebox{0.9\hsize}{!}{%
    $\D{\mathbf x_\tau} = \underbrace{\gamma(\mathbf y-\mathbf x_\tau)}_{:=\,\mathbf f(\mathbf x_\tau, \mathbf y)} \D{\tau}
        + \underbrace{\left[ \smin \left(\frac{\smax}{\smin}\right)^\tau \sqrt{2\log\left(\frac{\smax}{\smin}\right)} \right]}_{:=\,g(\tau)} \D{\mathbf w}$.
        }
\end{equation}
The stiffness hyperparameter $\gamma$ controls the slope of the decay from $\mathbf{y}$ to $\mathbf{x}_0$, and the noise extrema $\smin$ and $\smax$ control the noise scheduling, i.e. the amount of white Gaussian noise injected at each timestep during the forward process. 
Therefore, the forward process in Eq. \eqref{eq:ouve-sde}, 
injects an infinitesimal amount of corruption $\gamma (\mathbf y - \mathbf{x}_t )\D{\tau}$ to the current process state $\mathbf x_\tau$, along with Gaussian noise with standard deviation $g(\tau) \D{\tau}$. 
It is shown in \cite{Welker2022SGMSE} that the solution to \eqref{eq:ouve-sde} admits a complex Gaussian perturbation kernel $p(\mathbf{x}_\tau | \mathbf{x}_0, \mathbf{y})$ with
mean $\boldsymbol \mu(\mathbf{x}_0, \mathbf{y}, \tau)$ and variance $\sigma(\tau)^2$:
\begin{equation}
\label{eq:mean}
    \boldsymbol\mu(\mathbf x_0,\mathbf y, \tau) = \mathrm e^{-\gamma \tau} \mathbf x_0 + (1-\mathrm e^{-\gamma \tau}) \mathbf y
    \,,
\end{equation}%
\begin{equation}
    \label{eq:std}
    \sigma(\tau)^2 = \frac{
        \smin^2\left(\left(\sfrac{\smax}{\smin}\right)^{2\tau} - \mathrm e^{-2\gamma \tau}\right)\log(\sfrac{\smax}{\smin})
    }{\gamma+\log(\sfrac{\smax}{\smin})}
    \,.
\end{equation}
\vspace{-1em}

\subsection{Score function estimator}\label{sec:training}

During inference, the score function $\sco$ is not known and must be estimated by a so-called \emph{score model} $\mathbf{s}_\theta$.
Once obtained, all quantities are available for solving Eq. \eqref{eq:reverse-sde} with classical numerical methods (see Section~\ref{sec:inference}).  
Given the Gaussian form of the perturbation kernel $p(\mathbf x_\tau|\mathbf x_0, \mathbf y)$%
, the following \emph{denoising score matching} objective can be used to train the score model $\mathbf{s}_\phi$ \cite{vincent2011connection}:
\begin{equation}\label{eq:training-loss}
\resizebox{0.85\hsize}{!}{%
        $\mathcal{J}^{(\mathrm{DSM})}(\phi)
    = \mathbb{E}_{t,(\mathbf x_0,\mathbf y), \mathbf z, \mathbf x_\tau} \left[
        \norm{\mathbf s_\phi(\mathbf x_\tau, \mathbf y, \tau) + \frac{\mathbf z}{\sigma(\tau)}}_2^2
    \right].$}
\end{equation}
To optimize $\eqref{eq:training-loss}$, a clean utterance $\mathbf{x}_0$ and noisy utterance $\mathbf{y}$ are first picked in the training set. 
A diffusion time step $\tau$ is sampled uniformly in $[\tau_\epsilon, T]$ where $\tau_\epsilon > 0$ is a minimal diffusion time used to avoid numerical instabilities.
Then the current process state is obtained by Gaussianity of the perturbation kernel as $\mathbf x_\tau =  \boldsymbol\mu(\mathbf x_0,\mathbf y, \tau) + \sigma(\tau) \mathbf z$,  with $\mathbf z \sim \mathcal{N}_\mathbb{C}\left(\mathbf z; \mathbf 0, \mathbf{I}\right)$.
Classical gradient descent methods are then used to tune the score model (see Section~\ref{sec:exp:hyperparameters}).

\subsection{Inference through reverse sampling}\label{sec:inference}

At inference time, we sample $\mathbf x_T$, with:
\begin{equation}
    \mathbf x_T \sim \mathcal N_{\mathbb C}(\mathbf x_T; \mathbf y, \sigma^2(T) \mathbf I).
\end{equation}

Conditional generation is then performed by solving the reverse \ac{sde} \eqref{eq:reverse-sde} from $\tau=T$ to $\tau=0$, where the score function is replaced by its estimator $\mathbf s_\phi$.
We use 
classical SDE numerical solvers \cite{song2021sde} based on 
a discretization of 
\eqref{eq:reverse-sde} according to a uniform grid of $N$ points on the interval $[0, T]$ (no minimal diffusion time is needed here). 
We will denote by $G_\phi$ the generative model corresponding to 
reverse diffusion 
such that the clean speech estimate is $\widehat{\mathbf{x}} = \mathbf{x}_0 = G_\phi(\mathbf{y})$.

%% file: sections/storm.tex
\begin{figure*}
\centering
\includegraphics[width=0.99\textwidth]{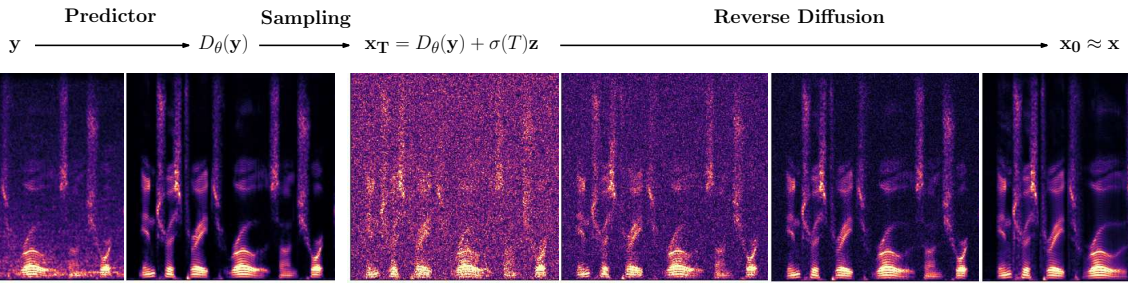}
\caption{\protect\centering StoRM inference process. The predictive stage produces a denoised version $D_\theta(\mathbf{y})$. Reverse diffusion $G_\phi$ is then carried out by first adding Gaussian noise $\sigma(T)\mathbf{z}$ to obtain the start sample $\mathbf{x}_T$, and finally by solving the reverse diffusion \ac{sde} \eqref{eq:reverse-sde} to obtain the estimated clean speech $\mathbf{x}_0$.}
\label{fig:stochastic-regeneration-inference}
\vspace{-1em}
\end{figure*}

We now revisit our \textbf{Sto}chastic \textbf{R}egeneration \textbf{M}odel \linebreak (StoRM) combining predictive and generative modelling originally proposed in \cite{Lemercier2022storm}.
An initial predictor $D_\theta$ is used as a first stage to generate a denoised version of the sample (see Figure~\ref{fig:stochastic-regeneration-inference}). This estimate can be polluted by residual noise and speech distortions due to the
the fact that predictive models trained with a mean-square error objective map noisy speech to the posterior mean $\mathbb{E}[\mathbf{x} | \mathbf{y}]$ 
rather than to a sample of the posterior distribution  
\cite{MurphyBook2, 
Lemercier2022storm
}.
A generative diffusion model $G_\phi$ then
learns to regenerate the clean speech $\mathbf{x}_0$  given 
$D_\theta(\mathbf{y})$:
\begin{equation}
    \hat{\mathbf{x}} = G_\phi( D_\theta( \mathbf{y} ) ).
\end{equation}
The inference process is shown in \figurename~\ref{fig:stochastic-regeneration-inference}.
For training, we use a criterion $\mathcal{J}^{(\mathrm{StoRM})}$ combining denoising score matching $\mathcal{J}^{(\mathrm{DSMS})}$ (where the difference with \eqref{eq:training-loss} is the presence of $D_\theta(\reverb)$ as extra-conditioning)
and a supervised regularization term $\mathcal{J}^{(\mathrm{Sup})}$ matching the output of the initial predictor to the target speech:
\begin{equation} \label{eq:regen-loss}
\begin{aligned}
    & \resizebox{0.85\hsize}{!}{
        $\mathcal{J}^{(\mathrm{DSMS})}(\theta) = \mathbb{E}_{\tau,(\mathbf x,\mathbf y), \mathbf z} 
        \norm{\mathbf s_\phi(\mathbf x_\tau, \left[ \mathbf y, D_\theta(\mathbf{y}) \right],  \tau) + \frac{\mathbf z}{\sigma(\tau)}}_2^2$},  \\[1pt]
    & \resizebox{0.55\hsize}{!}{
        $\mathcal{J}^{(\mathrm{Sup})}(\phi) =
    \mathbb{E}_{(\mathbf x,\mathbf y)} 
        \norm{\mathbf{x} - D_\theta(\mathbf{y}) }_2^2$},  \\[5pt]
    & \resizebox{0.7\hsize}{!}{
        $\mathcal{J}^{(\mathrm{StoRM})}(\theta, \phi) = \mathcal{J}^{(\mathrm{DSMS})}(\theta) + \alpha \mathcal{J}^{(\mathrm{Sup})}(\phi)$},
    \end{aligned}
\end{equation}
\noindent where a value of $\alpha=1$ is empirically chosen.
As $D_\theta(\mathbf{y})$ may not be a sufficient cue for optimal reconstruction of the target speech, we additionally provide $\mathbf y$ as conditioning to the score model $\mathbf{s}_\phi$ by stacking it with $D_\theta(\mathbf y)$ (see Section~\ref{sec:network}).

%% file: sections/exp.tex
\subsection{Data}
We generate our simulated dataset 
using clean speech data from the WSJ0 corpus and simulated and recorded wind noise, each making up for half of the noise data.
The simulated half of the noise dataset is created with the wind noise generator 
\cite{Mirabilii2022simulating}. Wind noise with airflow speed-dependent behaviour is generated using randomized airflow profiles (see Table~\ref{tab:data}).
The real-recorded other half of the noise dataset is obtained from public sources such as Freesounds (4.3 h), YouTube (0.1 h) and various open-source noise databases (1.8 h)\cite{IKSWindNoiseData, WindNoiseDataset2022, TestData2020}.

We design a non-additive speech in noise model by taking into account both non-linearities caused by microphone membrane displacement and clipping in case of strong wind. 
First, wind noise and speech signals are mixed additively with a random SNR.
Then, the membrane displacement non-linearity is simulated by using a compressor on the speech signal, sidechained by the noise signal. If the wind noise signal exceeds the compressor threshold, the speech signal is compressed by an amount determined by the compressor ratio and the magnitude of the noise signal above the compressor threshold. We sample compressor threshold, ratio, attack and release parameters to mimic various recording devices. Finally, hard-clipping is simulated by limiting the dynamic range of the noisy signal $\reverb$ between $-\eta \max{}(|\reverb|) $ and $\eta  \max{}(|\reverb|)$. We refer the reader to Table \ref{tab:data} for the data generation parameters.
In total 25, 2.3 and 1.5 hours of noisy speech sampled at 16kHz are created for training, validation and testing respectively. We make our data generation method publicly available\footnote{https://github.com/sp-uhh/storm}.

Finally, we also use an unseen dataset using real wind noise recorded in a wind tunnel, added to German speech with a SNR in $\{ 0, -5, -10 \} \, \mathrm{dB}$. For this data, provided by Advanced Bionics, only noisy speech without ground truth is available.

\input{tables/parameters}

\subsection{Hyperparameters and training setting}
\label{sec:exp:hyperparameters}

\paragraph{Data representation}
Noisy and clean utterances are transformed using a \ac{stft} with a window size of 510, a hop length of 128 and a square-root Hann window, at a sampling rate of $16$kHz, as in \cite{Richter2022SGMSE++,Lemercier2022storm}. A square-root magnitude warping is used to reduce the dynamical range of spectrograms \cite{Richter2022SGMSE++}.
During training, sequences of 256 \ac{stft} frames ($\approx$2s) are  extracted from the full-length utterances with random offsets and normalized by the maximum absolute value of the noisy utterance. 

\paragraph{Forward and reverse diffusion}
For the proposed stochastic regeneration model, we fix the stiffness to $\gamma=$1.5, the extremal noise levels to $\sigma_\mathrm{min}=$~0.05 and $\sigma_\mathrm{max}=$~0.5, and the extremal diffusion times to $T=1$ and $\tau_\epsilon=$~0.03 as in \cite{Lemercier2022storm}.
$N=20$ time steps are used for reverse diffusion using the first-order Euler-Maruyama prediction scheme, resulting in 21 neural network calls.

\paragraph{Network architecture} \label{sec:network}

For score estimation and initial prediction, we use two \linebreak copies of a lighter configuration of the NCSN++ architecture \cite{song2021sde}, which was proposed in our previous study \cite{Lemercier2022analysing} and denoted as \textit{NCSN++M} and has roughly 27.8M parameters.
For initial prediction, the noisy speech spectrogram $\reverb$ real and imaginary channels are stacked and provided as sole input to the network $D_\theta$, and no noise-conditioning is used.
For score estimation during reverse diffusion, the noisy speech spectrogram $\reverb$, the initial prediction $D_\theta(\reverb)$ and the current 
estimate $\state$ real and imaginary channels are stacked and fed to the network $\mathbf{s}_\phi$, and the current noise level $\sigma(\tau)$ is provided as a conditioner. The resulting approach is denoted as \textit{StoRM}.

We also investigate using GaGNet for initial prediction \cite{Li2022GaGNet}, a state-of-the-art predictive denoising approach conducting parallel magnitude- and complex-domain enhancement in the \ac{tf} domain. We use 257 frequency bins instead of the original 161 for compatibility with NCSN++-based score estimation, increasing the network capacity to 11.6M parameters compared to the original 5.9M. The resulting approach is denoted as \textit{StoRM-G}.

\paragraph{Baselines}

We compare our approaches to the purely generative \linebreak SGMSE+M \cite{Richter2022SGMSE++} and purely predictive NCSN++M \cite{Lemercier2022analysing}.
SGMSE+M uses the NCSN++M architecture for score estimation, $N=30$ reverse time steps with a Euler-Maruyama predictor and one step of Annealed Langevin Dynamics correction with step size $r=0.5$, resulting in 60 neural network calls.
We change the stiffness to $\gamma=2.5$ and maximal noise level to $\smax=0.75$.
We noticed that higher maximal noise level and stiffness were needed, as the initial mean $\reverb$ which needs masking by the Gaussian noise $\sigma(T)\mathbf{z}$ has higher energy compared to StoRM where the initial mean is $D_\theta(\reverb)$.

We also report the performance of the soft audio noise masking model using fully connected networks \linebreak (\textit{FCN+SANM}) \cite{Bai2018DNNWind} and the "Unified" version of the deep bidirectional long-short term memory network approach (\textit{DBLSTM-U}) by \cite{Lee2017DeepBL}, which is the state-of-the-art DNN-based method for wind noise reduction.

\paragraph{Training configuration}
We train the approaches NCSN++M, SGMSE+M, StoRM and StoRM-G using the Adam optimizer \cite{kingma2015adam} with a learning rate of $0.0005$ and an effective batch size of 16. We track an exponential moving average of the DNN weights with a decay of 0.999
\cite{Song2020Improved}. We train \acp{dnn} for a maximum of 500 epochs using early stopping based on the validation loss with a patience of 10 epochs.
For StoRM approaches, the initial predictor is pre-trained with a complex spectrogram mean-square error loss, then we jointly train the predictor and score network with \eqref{eq:regen-loss} \cite{Lemercier2022storm}. 
We implement FCN+SANM and DBLSTM-U using the hyperparameters and training configuration proposed by the authors.

\subsection{Evaluation metrics}
\label{sec:exp:eval}

For instrumental evaluation of the speech enhancement and dereverberation performance with clean test data available, we use intrusive measures such as \ac{pesq} \cite{Rix2001PESQ} to assess speech quality, \ac{estoi} \cite{Jensen2016ESTOI} for intelligibility and \ac{sisdr}
\cite{Leroux2019SISDR} for wind noise and distortion removal.
For reference-free assessment of speech restoration, we also use the non-intrusive DNSMOS
\cite{reddy2021dnsmos} and WVMOS
\cite{Andreev2022Hifi++} metrics, which perform DNN-based mean opinion \linebreak score estimation.
\input{tables/simulated.tex}

\input{tables/real.tex}

%% file: tables/parameters.tex
\begin{table}[t]
    \centering
    \begin{tabular}{c|c|c}
        \toprule \midrule
         \textbf{Parameter} & \textbf{Unit} & \textbf{Distribution} \\
         \midrule
         Number of wind gusts & & $\mathcal{U}(1, 10)$ \\
         Input SNR & dB & $\mathcal{U}(-6, 14)$ \\
         Compressor ratio & & $\mathcal{U}(1, 20)$ \\
         Compressor sidechain input level & & $\mathcal{U}(0.8, 1.2)$ \\
         Compressor attack & ms & $\mathcal{U}(5, 100)$ \\
         Compressor release & ms & $\mathcal{U}(5, 500)$ \\
         Clipping presence & & $\mathcal{B}(0.75)$ \\
         Clipping threshold $\eta$ & & $\mathcal{U}(0.85, 1.0)$ \\
         \midrule \bottomrule
         
    \end{tabular}
    \caption{\protect\centering Data generation parameters}
    \vspace{-1.5em}
    \label{tab:data}
\end{table}

%% file: tables/simulated.tex
\begin{table*}[t]
    \centering
    \begin{tabular}{c|c|ccccc}
    \toprule \midrule
         Method & \#Params & DNSMOS & WVMOS 
         & PESQ & ESTOI & SI-SDR  \\
         \midrule
         Noisy & & 3.04 $\pm$ 0.61 & 1.24 $\pm$ 2.58 
         & 1.70 $\pm$ 0.61 & 0.76 $\pm$ 0.19 & 4.1 $\pm$ 5.9 \\ \midrule
         
         FCN+SANM \cite{Bai2018DNNWind} & 4.3 M & 2.63 $\pm$ 0.66 & 2.17 $\pm$ 1.73 
         & 2.01 $\pm$ 0.57 & 0.78 $\pm$ 0.15 & 9.0 $\pm$ 4.3 \\ 
         
         DBLSTM-U \cite{Lee2017DeepBL} & 73.5 M & 3.50 $\pm$ 0.72 & 3.61 $\pm$ 0.49 
         & 2.94 $\pm$ 0.78 & 0.90 $\pm$ 0.10 & 15.5 $\pm$ 6.5 \\

         NCSN++M \cite{Lemercier2022analysing} & 27.8 M & 4.09 $\pm$ 0.39 & 3.70 $\pm$ 0.53 
         & 2.76 $\pm$ 0.92 & \textbf{0.92 $\pm$ 0.08} & \textbf{18.8 $\pm$ 6.2} \\
         
         SGMSE+M \cite{Richter2022SGMSE++} & 27.8 M & 4.01 $\pm$ 0.32 & 3.79 $\pm$ 0.40 
         & 2.83 $\pm$ 0.78 & 0.90 $\pm$ 0.10 & 16.5 $\pm$ 6.1 \\ \midrule
         
         StoRM (prop.) & 56.0 M & \textbf{4.19 $\pm$ 0.30} & 3.80 $\pm$ 0.43 
         & 3.02 $\pm$ 0.76 & 0.91 $\pm$ 0.08 & 17.4 $\pm$ 6.0 \\
         
         StoRM-G (prop.) & 39.6 M & \textbf{4.19 $\pm$ 0.30} & \textbf{3.87 $\pm$ 0.41} 
         & \textbf{3.07 $\pm$ 0.76} & \textbf{0.92 $\pm$ 0.08} & 17.6 $\pm$ 6.0 \\

         \midrule \bottomrule
    \end{tabular}
    \caption{\protect\centering Enhancement results on our simulated test set.
    Values indicate mean and standard deviation.
    }
    \label{tab:results}
    \vspace{-1em}
\end{table*}

%% file: tables/real.tex
\begin{table}[t]
    \centering
\scalebox{1}{
    \begin{tabular}{c|cc}
    \toprule \midrule
         Method & 
         DNSMOS & WVMOS \\
         \midrule
         Noisy& 1.89 $\pm$ 0.41 & 0.08 $\pm$ 0.19 \\ \midrule
         
         FCN+SANM \cite{Bai2018DNNWind} & 1.29 $\pm$ 0.33 & 0.23 $\pm$ 0.34 \\
         
         DBLSTM-U \cite{Lee2017DeepBL}& 1.96 $\pm$ 0.47 & 0.23 $\pm$ 0.33 \\

         NCSN++M \cite{Lemercier2022analysing} & 3.34 $\pm$ 0.59 & 1.59 $\pm$ 0.55 \\

         SGMSE+M \cite{Richter2022SGMSE++} & 3.44 $\pm$ 0.11 & 1.52 $\pm$ 0.50 \\  \midrule
         
         StoRM (prop.) & 3.36 $\pm$ 0.44 & 1.33 $\pm$ 0.60 \\
         
         StoRM-G (prop.) & \textbf{3.56 $\pm$ 0.42} & \textbf{1.67 $\pm$ 0.57} \\
         \midrule \bottomrule
    \end{tabular}
    }
    \caption{\protect\centering Enhancement results on the unseen dataset using real-recorded wind noise. Values indicate mean and standard deviation.
    }
    \label{tab:results-real}
    \vspace{-1em}
\end{table}

%% file: sections/results.tex
\subsection{Simulated dataset}

We report in Table~\ref{tab:results} instrumental metrics for the proposed method and baselines on the proposed simulated 
test set. 
We observe that the FCN+SANM baseline \cite{Bai2018DNNWind} hardly improves 
over noisy speech, as it uses a simplistic low-\linebreak capacity architecture without any sequence-modelling module. In comparison, DBLSTM-U \cite{Lee2017DeepBL} yields good results for a simple predictive approach but has a large number of parameters.
As already reported in \cite{Lemercier2022analysing, Lemercier2022storm}, predictive \linebreak NCSN++M yields high ESTOI and SI-SDR but mediocre quality-related metrics, due to important speech distortions. Purely generative SGMSE+M achieves marginally higher PESQ and WVMOS but lower ESTOI and SI-SDR, and produces many generative artifacts.

The proposed methods StoRM and StoRM-G highly improve speech quality, while remaining competitive with NCSN++M in terms of ESTOI and SI-SDR and using three times fewer operations than SGMSE+M. 
StoRM-G slightly outperforms StoRM with fewer parameters, showing the efficiency of using GaGNet as initial predictor.

\subsection{Real-recorded dataset}

We display in Table~\ref{tab:results-real} instrumental metrics of the different baselines and proposed models on the unseen dataset using real-recorded wind noise. 
We show that NCSN++M generalizes well to unseen noisy data for a predictive approach, compared to the other predicitve baselines. However, SGMSE+M and StoRM-G perform much better, the latter improving DNSMOS by 1.8 points.

%% file: sections/conclusion.tex
We propose to solve the wind noise reduction task with our previously proposed diffusion-based stochastic regeneration model, combining predictive and generative modelling. We design a speech in noise signal model which deviates from the classical additive model by introducing non-linearities to simulate membrane displacement and clipping.
We show that the introduced method is able to strongly increase the quality and intelligibility of speech in wind noise. The proposed stochastic regeneration model outperforms previous DNN-based methods for wind noise reduction as well as purely predictive and generative methods in terms of instrumental metrics.
In particular, it generalizes well to unseen data using real-recorded wind noise.